\documentclass[%
reprint,
superscriptaddress,
amsmath,amssymb,D_1
aps,
]{revtex4-2}
\usepackage{graphicx}
\usepackage{dcolumn}
\usepackage{bm}
\usepackage{mathrsfs}
\usepackage{amsmath}
\usepackage[usenames,dvipsnames]{color} 
\usepackage{amssymb}
\usepackage{slashed}

\usepackage{subfigure} 
\usepackage{color,soul} 
\setul{0.5ex}{0.3ex}
\setulcolor{red}
\setstcolor{red}

\usepackage{hyperref}
\usepackage[mathlines]{lineno}

\begin{document}
	
	\title{Dark Matter Jets of Rotating Black Holes}
	
	\author{Ottavia Balducci}
	\email{ottavia.balducci@physik.uni-muenchen.de}
	\affiliation{Arnold Sommerfeld Center for Theoretical Physics, Theresienstra{\ss}e 37, 80333 M\"unchen\\}
	
	\author{Stefan Hofmann}
	\email{stefan.hofmann@physik.uni-muenchen.de}
	\affiliation{Arnold Sommerfeld Center for Theoretical Physics, Theresienstra{\ss}e 37, 80333 M\"unchen\\}

	\author{Maximilian Koegler}
	\email{m.koegler@physik.uni-muenchen.de}
	\affiliation{Arnold Sommerfeld Center for Theoretical Physics, Theresienstra{\ss}e 37, 80333 M\"unchen\\}
	
	\date{\today}
	
	\begin{abstract}
		We present a novel approach which produces Dark Matter
		jets along the rotation axis of Kerr black holes
		utilizing the Penrose process. The properties of these
		jets are investigated, as well as their potential to
		create Dark Matter overdensities in the solar system
		and in the vicinity of the black hole. We discover a
		highly collimated and long-range Dark Matter jet with
		a density that is most sensitive to
		the mass and distance to the black hole.
	\end{abstract}
	
	\maketitle
	
	\section{Introduction}
	
	Astrophysical jets produced by supermassive black holes are
	usually explained by the magnetic Blandford-Znajek process
	\cite{Blandford:1977ds}. Due to its electromagnetic nature,
	only charged particles can be ejected as a jet in this way. In
	this work, we quantitatively analyze a gravitational
	mechanism for jet production. Through the Penrose process
	\cite{Penrose:1971uk}, particles which have fallen into a
	supermassive black hole's ergosphere from the accretion disk
	can extract energy from the black hole and be
	ejected along the rotation axis \cite{Gariel:2010}. It should
	be noted that while this is subdominant with respect to the Blandford-Znajek mechanism, it can affect Dark Matter (DM)
	particles as well, due to its gravitational nature.
	
	The main goal of this article is to confirm the presence of a
	DM beam produced by the mechanism described above
	quantitatively. It will be shown that the DM beam is highly
	collimated. Furthermore, the dependence of the DM density on the
	mass of the black hole and on the distance to it will be
	analyzed as well.
	
	The existence of such a DM beam renders this phenomenon
	potentially interesting for future DM
	detection. Indeed, a number of recent tests have yielded
	promising results for indirect DM detection: records from the PAMELA satellite
	\cite{Adriani:2008zr} suggest a surprisingly high fraction of
	positrons in cosmic ray measurements over {$10\,$}GeV. Similar
	results were obtained by the AMS collaboration
	\cite{Aguilar:2013qda}, the DAMPE collaboration
	\cite{Ambrosi:2017wek} reported a peak in the flux of electron
	and positron cosmic rays at around $1.4\,$TeV, the
	H.E.S.S. telescopes \cite{Aharonian:2004wa} detected a
	point-like source of very high-energy $\gamma$ rays from
	Sagittarius A*, and the FERMI/LAT data hinted at an
	excess in {$\gamma$} rays at energies of {$130\,$}GeV
	\cite{Weniger:2012tx}.  Of course, it is unknown whether the
	detected signals are truly the result of DM particle
	annihilation, but it is an interesting possibility.
	
	The usual WIMP cross section is too small to explain the
	observed fluxes of DM-annihilation products, so a boost factor
	$B$, which usually lies in the range between {$10^{2}$}
	and {$10^4$}, must be artificially added to DM models in order
	to explain the observed measurements,
	\cite{Bergstrom:2005ss,Bergstrom:2004cy,Hooper:2008kv,Meade:2009iu,Bergstrom:2009fa}. Several
	potential possibilities for the physical origin of this
	overdensity have been discussed in the literature such as
	density inhomogeneities at tiny scales \cite{Lavalle:2007apj}
	and Sommerfeld enhanced annihilation cross sections
	\cite{Hisano:2003ec}.  When an incoming beam of DM particles
	annihilates with some target DM particles in the proximity of
	the Earth, the boost factor $B$ is defined as
	
	\begin{equation}
		B=\frac{\rho_{B}}{\rho_{0}}\frac{\rho_{T}}{\rho_{0}}
	\end{equation}
	where {$\rho_{B}$} and {$\rho_{T}$} are the densities of the
	beam and the target, respectively, and {$\rho_{0} \approx
		0.4\,$GeV/cm$^3$} is the macroscopic DM energy density in
	the solar system, as defined in \cite{Bergstrom:2009ib} and
	\cite{Bergstrom:2012fi}.
	
	With this motivation in mind, we study whether such DM
	overdensities in our vicinity could be caused by rotating
	black holes via jet production.  In this situation, the target
	DM particles are those in our immediate vicinity, therefore
	{$\rho_{T}/\rho_{0}=1$}.  We will show that even though this
	mechanism produces a DM beam, the corresponding DM densities
	are not sufficient to obtain a significant boost factor.
	
	This article is organized as follows: we describe geodesics in
	a spinning black hole geometry in Sec.~\ref{sec:geometry} and
	gather the relevant ones in Sec.~\ref{sec:boost} to determine
	the overdensities in the black hole jet.  We introduce
	approximations in Sec.~\ref{sec:approximation}, which allow us
	to numerically evaluate the DM density in the beam and show
	the findings in Sec.~\ref{sec:applications}. We conclude in
	Sec.~\ref{sec:conclusion}. Throughout this article, we use
	units where $G=c=1$.
	
	\section{Geodesics in a Kerr Geometry}
	
	\label{sec:geometry}
\begin{figure}[t]
		\centering
		\includegraphics[scale=1]{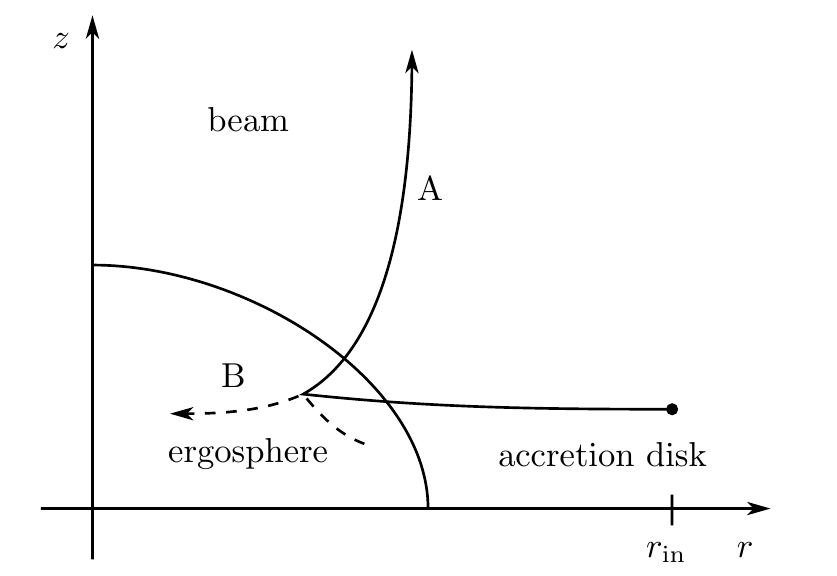}
		\caption{Relevant geodesics for
			jet formation in a Kerr spacetime. The solid line depicts a geodesic of
			the particle A in the accretion disk
			falling into the black hole where it scatters off
			the particle B following the dashed line. After the
			collision, particle A leaves the black hole and moves
			parallel to the rotation axis and particle B falls
			towards the black hole
			singularity.}
		\label{boost}
	\end{figure}

	The system under consideration is a galaxy with a rotating
	Kerr black hole in the center. We concentrate on particles
	which follow geodesics leading them from the accretion disk
	into the ergosphere. There, they can scatter with other
	particles or simply decay. As depicted in Fig.~\ref{boost},
	some of the products of these interactions can then follow a
	geodesic leading out of the ergosphere and moving parallel to
	the rotation axis due to the Penrose process as shown in
	Ref.~\cite{Gariel:2010}. Collecting all geodesics with similar
	behaviour results in the black hole agglomerating particles
	from the accretion disk and releasing them along its rotation
	axis and thus forming a jet.
	
	We consider the usual Kerr metric of a black hole with mass $M$ and
	angular momentum $M a$.  The motion of a particle for a non-rotating
	neutral black hole ($a=0$) is completely determined by the particle’s
	mass, energy and angular momentum. However, for the case of a Kerr
	black hole ($a \not = 0$) a fourth constant is needed, the Carter
	constant, which for a particle of unit mass reads
	\begin{equation}
		Q=\left(u_{\theta}\right)^{2}+a^{2}\cos^{2}\left(\theta\right)\left(1-E^{2}\right)+\frac{\cos^{2}\left(\theta\right)}{\sin^{2}\left(\theta\right)}L^{2}\;{}\label{carter-constant-mass}
	\end{equation}
	where $u_\theta=\mathrm{d}x_{\theta}/\mathrm{d}\tau{}$ is the {$\theta$}
	component of the proper four-velocity, $E=-u_{t}$ is the
	particle's energy and $L=u_{\phi}$ is its angular momentum. With the help of the constants of motion
	together with the initial position and velocity of the DM particle in
	the accretion disk we can determine whether it has a chance
	of ending up in the DM
	jet together with its final position and velocity for a given time. In
	the next section we take all these geodesics into account to obtain the DM
	density in the jet.

	\section{The DM density of the
		ejected particles}\label{sec:boost}
	
	To investigate the jet formation we obtain the overall density of the outgoing particles by taking
	into account the contributions due to all the possible outgoing
	velocities,
	\begin{equation}
		\rho_{\rm{}out}\left(r_{\mathrm{out}},\theta\right)=\int{} \frac{\mathrm{d}^{3}\mathbf {v}_{\rm{}out}}{r_{\mathrm{out}}^{2}\sin\left(\theta\right)v_{r,\,\mathrm{out}}}{\frac{\mathrm{d}^{3}N_{\rm{}out}\left(r_{\mathrm{out},\theta,\mathbf{v}_{\mathrm{out}}}\right)}{\mathrm{d}{}v_{r,{\rm{}out}}\mathrm{d}{}v_{\phi,{\rm{}out}}\mathrm{d}{}v_{\theta,{\rm{}out}}}}\;{},\label{rho-out}
	\end{equation}
	where
	{$N_{\rm{}out}\left(r_{\mathrm{out}},\theta{},\mathbf{v}_{\rm{}out}\right)$}
	is the total number of particles in the beam at
	{$\left(r_{\mathrm{out},\theta}\right)$} with velocity
	{$\mathbf{v}_{\rm{}out}$}. The number of outgoing particles is
	computed by taking the number of particles that fall into the
	ergosphere and filtering out all of those that follow
	trajectories that do not end up in the beam,
	\begin{equation}
		N_{\rm{}out}\left(\mathbf{x},\mathbf{v}_{\rm{}out}\right)=\eta P_\text{scat}\int{}\mathcal{D}^{3}\mathbf{v}_\text{in}\rho_{\rm{}in} V_\text{in}\left(\mathbf x, \mathbf  v_{\rm{}in}, \mathbf v_\text{out}\right)\,{}.\label{nout}
	\end{equation}
	Here $V_\text{in}$ is the volume occupied by the infalling
	particles with initial velocities
	{$\mathbf{v}_\text{in}=\left(v_{r},v_{\phi},v_{\theta}\right)$}
	which end up in the beam and reach $\mathbf x$ with final
	velocity $\mathbf v_\text{out}$. {$\eta$} is the efficiency of
	the Penrose process. $P_\text{scat}$ gives the probability of
	an infalling particle scattering in the ergosphere and not
	before. The density of the infalling particles is denoted by
	{$\rho_{\rm{}in}$}. The integration runs over
	{$\mathcal{D}^{3}\mathbf{v}=\mathrm
		d^{3}\mathbf{v}f\left(\mathbf{v}\right)$}, where
	{$f\left(\mathbf{v}\right)$} is the distribution function for
	the velocities of the infalling particles. In the next
	Section, we will introduce approximations to simplify
	Eqs.~(\ref{rho-out}) and (\ref{nout}), and afterwards compute
	{$\rho_{\mathrm{out}}$} numerically.
	
	\section{Approximations and Assumptions}\label{sec:approximation}
	
	Since the system under consideration has a high dimensional parameter
	space and partially includes unknown initial conditions,
	approximations and assumptions are needed to
	numerically compute Eqs.~(\ref{rho-out}) and (\ref{nout}) for
	any specific galaxy. This section
	provides a discussion of all such approximations and assumptions used in this
	article.
	
	\textbf{Volume of infalling particles:}
	
	The volume occupied by the incoming particles is for convenience approximated 
	to be a ring of radius {$r_{\rm{}in}$}, height
	{$\Delta{}z=z_{\rm{}max}-z_{\rm{}min}$}, and width
	{$v_{r}\Delta{}t$}, i.e. ${V_\text{in}\left(\mathbf x, \mathbf v_{\rm{}in}, \mathbf
	v_{\rm{}out}\right) =
	2\pi{}r_{\rm{}in}v_{\rm{}r}\Delta{}t\Delta{}z\left(\mathbf x,
	\mathbf v_{\rm{}in}, \mathbf v_{\rm{}out}\right)}$. A more sophisticated procedure would be to
	start from the detection of the particles with given
	velocities in some time interval and trace them back to the
	ergosphere and to the accretion disk. In this way one would
	collect all relevant geodesics such as the one in
	Fig.~\ref{boost} and could ask where the detected particles
	originated in the accretion disk, which DM densities were
	present at that time and thus how many particles are actually
	taking this geodesic. By approximating this with an infalling
	ring, there are different counteracting effects, e.g. a larger
	radius accounts for a bigger volume but also a smaller
	density, there is a region in the accretion disk in which the
	most particles relevant for jet formation
	originate. Therefore, a suitable choice of $r_\text{in}$ has
	to lie within this region in order to gather the most relevant
	geodesics and thus to be a reasonable approximation to the
	above procedure. To obtain an upper bound, we choose
	$r_\text{in}$ so that the maximum feasible
	$\rho_{\mathrm{out}}$ is achieved. For example, for the
	Andromeda galaxy {the correct choice is {$r_{\mathrm{in}}=$}}
	$0.1\,$pc.
	
	\textbf{Carter constant:}
	
	We assume that the value of the Carter constant does not
	change due to the scattering in the ergosphere, as was done in
	Ref.~\cite{Gariel:2010}. Furthermore, we assume that
	outgoing particles with the correct Carter constant
	{$Q_{\rm{}out}$} do indeed end up in the beam and do not
	follow some other geodesic with the same value of
	{$Q_{\rm{}out}$}. This leads to an upper bound for the
	maximal obtainable overdensity. Keeping all other
	parameters fixed and solving {$Q_{\rm{}in}=Q_{\rm{}out}$}
	for {$z\left(\mathbf x,
		\mathbf{v}_{\rm{}in},\mathbf{v}_{\rm{}out}\right)$}
	results in the position the particles initially must have
	in order to end up in the beam at $\mathbf x$. Here,
		{$Q_{\mathrm{in}}$} is the Carter constant for the
		ingoing particles. If it has no solution, there is no
	geodesic connecting the point in the beam with the point
	in the accretion disk.
	This
	makes sure that all and only the particles that can end up in the beam
	are considered for each set of parameters
	{$\{\mathbf x, \mathbf  v_{\rm{}in}, \mathbf  v_{\rm{}out}\}$}.
	
	\textbf{Distribution function:}
	
	The velocity distributions of the infalling particles are assumed to be
	Gaussian, 
	i.e. 
	\begin{equation}
		f\left(\mathbf{v}_\text{in}\right)\equiv{}\frac{\exp\left\{-\frac{1}{2} (\mathbf v_\text{in} - \mathbf v_0)^T \Sigma^{-1} (\mathbf v_\text{in} - \mathbf v_0)\right\}}{\sqrt{(2\pi)^{3}|\text{det}(\Sigma)|}} \,.
	\end{equation}
	We assumed the covariance matrix to be diagonal and isotropic
	such that it reads $\Sigma= \text{diag}(\sigma^2, \sigma^2,
	\sigma^2)$ with standard deviation $\sigma$. The orbiting of
	the black hole dominates the mean velocity of the DM particles
	in the halo, and consequently {$\mathbf{v_0} =
		(0,v_{\phi,0},0)$}. We take explicit values for $v_{\phi,
		0}$ from data on the rotation curve of the Milky Way in
	\cite{Boshkayev:2018sbj} and assume that other galaxies
		have similar velocity distributions. In order to estimate
	the value of {$\sigma$} we refer to Fig.~2 of
	Ref.~\cite{Brandt:2010ts} to obtain the mass accretion rate
	{$\mathrm dM/\mathrm dt$} of supermassive black holes. By
	setting this equal to the infall rate of particles around the
	black hole we obtain a value for {$\sigma$}. Explicitly, the
	equation to be solved is
	\begin{equation}
		\frac{\mathrm dM}{\mathrm dt}=4\pi{}r_{\rm{}in}^{2}\rho_{\rm{}in} \int_{0}^{1} \mathrm dv_{r} \frac{1}{\sqrt{2\pi{}\sigma^{2}}}e^{-\frac{v_{r}^{2}}{2\sigma^{2}}}v_{r} \label{dm-dt}\,{}.
	\end{equation}
	We solve Eq.~(\ref{dm-dt}) for {$\sigma$} numerically with the
	appropriate parameters depending on the specific case
	considered. The velocity distribution must then be multiplied
	with the density of DM at {$r_{\rm{}in}$} in order to obtain
	the total number of particles with velocity
	{$\mathbf{v}_{\rm{}in}$} per unit of volume. As an example, for the Andromeda galaxy with
	{$r_{\rm{}in}=0.1\,{}$}pc and assuming a cored DM profile
	\cite{Tulin:2017ara}, the DM density can be approximated by
	{$\rho_{\rm{}in}\left(0.1\,{\rm{}pc}\right)=30\,\rho_0$}.
	
	\textbf{Penrose efficiency:}
	
	The efficiency of the Penrose process depends on the angular momentum
	of the black hole and can reach for maximally rotating black holes a value of
	$\eta = 0.29$ \cite{Dolan:2011xt}. Since the angular momentum of the black
	holes under consideration is not known we assume a Penrose efficiency
	$\eta = 0.01$. Since the DM density depends linearly on $\eta$, one can easily adjust it for other angular momenta.
	
	\textbf{Mean free path:}
	
	In order to describe the path of incoming particles solely with one geodesic, the probability for avoiding any scattering in the accretion disk has to be implemented. Taking the mean free path of the particles such that in average one scattering event occurs within the ergosphere and assuming no significant change in DM density along the geodesic we take $P_\text{scat} = {\lambda_\text{mfp}}/{r_\text{in}} \approx {2M}/{r_\text{in}}$.
	
	\section{Results}\label{sec:applications}
	With the approximations in place, we determine the DM density
	in a black hole DM beam numerically, i.e.
	$\rho_{\mathrm{out}}(r_\text{out}, \theta)$ in
	Eq.~\eqref{rho-out}. To this end we integrate over the initial
	velocities of the DM particles in Eq.~\eqref{nout} as a
	Riemann sum. In order to reduce the computation time as much
	as possible the integration limits were chosen such that the
	neglected portion of the parameter space contributes at most
	{$0.1\%$} to the final result. Furthermore, when deciding on
	the step size for the integration, a compromise had to be
	taken.  However, while the final result can indeed change by
	as much as two orders of magnitude when decreasing the size of
	the integration steps, the qualitative analysis remains
	unchanged. 
	
	In the code provided in Ref. \cite{code} we implemented
		the steps just described. The integration limits and
	step size are numerically adjusted such that we use the sweet
	spot between accuracy and computation time required for each
	specific black hole in question. Additionally, for the
	outgoing radial velocity there is a lower bound since
	particles arriving to us today had to be sent by the black
	hole at the earliest during the black hole formation such that
	$v_{r,\text{out}}>r_\text{out}/t_\text{age}$, where
	$t_\text{age}$ is the age of the black hole. DM particles that
	are slower than this bound might reach us in the future, but
	cannot have reached us yet and thus do not contribute to the
	boost factor.
	
\begin{figure}[t]
		\centering
		\includegraphics[scale=1]{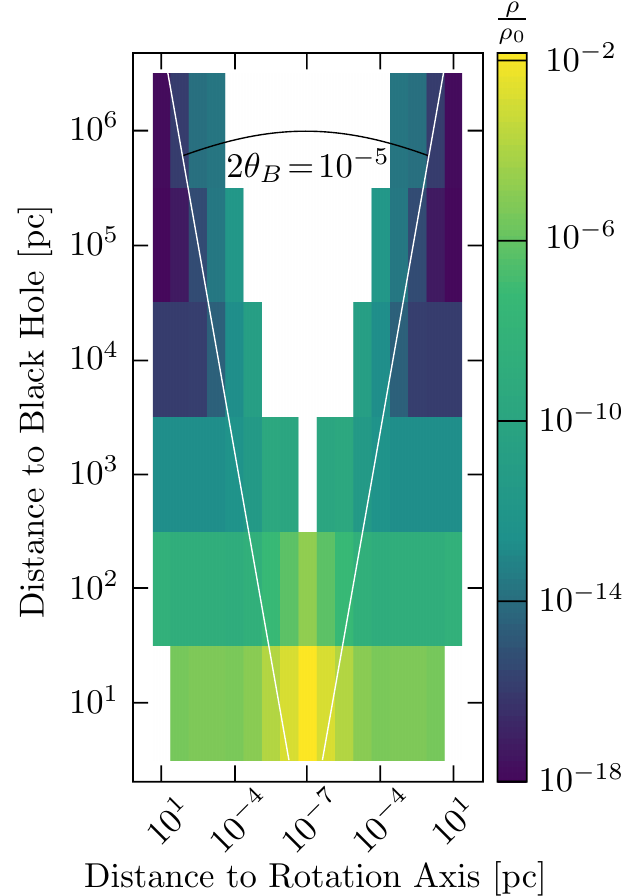}
		\caption{DM density profile in the beam represented in units of $\rho_{0}$ for various distances to the black hole $r_\text{out}$ and the rotation axis. In addition, the opening angle of the beam $2\theta_B$ is schematically indicated. Due to the inapplicability of our approximations, for the white pixels there is no data.
		}
		\label{boostAndr}
	\end{figure}

	In Fig.~\ref{boostAndr} we show the density profile of the DM
	beam created by the Andromeda black hole. There are immediate
	lessons to be learned from this result. First, the beam is
	highly collimated with an opening angle of $2 \theta_B \approx
	10^{-5}$ which for visibility is stretched. Second, the beam
	is far ranging such that even for a distance of $1\,$Mpc the
	beam can still be distinguished. Third, the DM density of the beam increases by
	many orders of magnitudes if evaluated closer to the black
	hole and its rotation axis but never reaches a significant
	overdensity. With these observations we can conclude that the
	Andromeda black hole is in principle capable to produce a
	sharp, far ranging but faint DM beam. As a sanity check we
	analyzed the qualitative behavior of the DM beam for different
	values of the black hole's angular momentum $a$. For a
	Schwarzschild black hole, i.e. $a=0$, without adjusting $\eta$, the DM density is 8
	magnitudes smaller compared to the case with $a=M/2$
	\cite{code}. This background contribution is due to geodesics
	reaching the target location without using any
	Kerr-black-hole-related effects such as the Penrose
	process. This contribution is negligible compared to the DM
	density in the beam and thus effects uniquely attributed to
	the Kerr metric are causing the DM beam. Furthermore, the
	larger $a$ is chosen, the more collimated is the DM beam as
	expected \cite{code}.
	
	It should be noted that the obtained densities are much smaller than
	the local DM density. As an example, we present here the
	explicit result for the black hole in the center of the Andromeda
	galaxy, for which we assume that its rotation axis points at
	us. Taking the age of the galaxy equal to the one of the black hole to
	be $t_\text{age} = 10^{10}\,\mathrm{yrs}$, the DM density as in
	Eq.~\eqref{rho-out} in our local neighborhood is
	$\rho_{\mathrm{out}}\approx{}10^{-12}\rho_{0}$. Despite the Andromeda
	black hole being the most promising candidate, since $B \ll 1$ the
	overdensity is negligible and cannot explain boost functions of the
	order $10^2$ required for explaining indirect DM measurements in the
	solar system.
	
	\begin{figure}[t]
		\centering
		\includegraphics[scale=1]{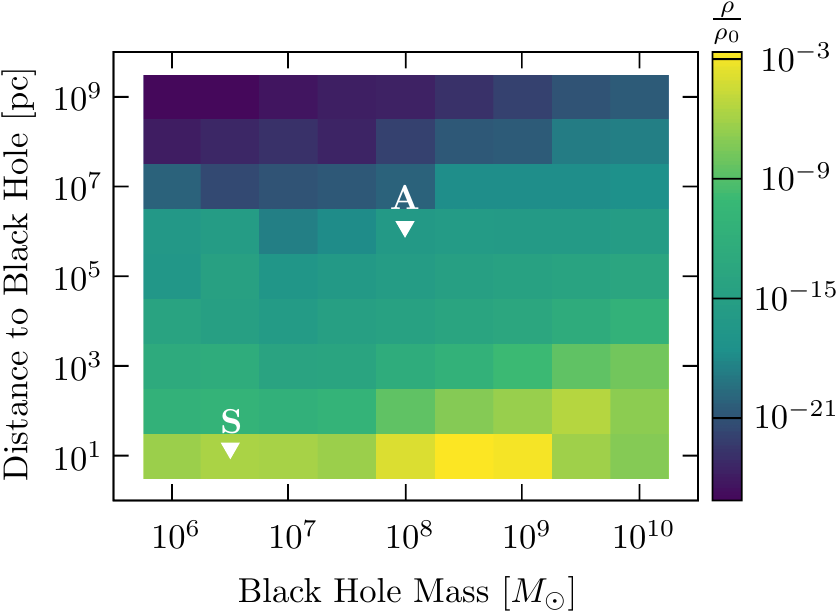}
		\caption{The DM
				density in the beam at various distances from a black hole $r_\text{out}$ with mass
			$M$ for $\theta=10^{-8}$. The mass range includes both smaller black
			holes at galactic centers and the largest black
			holes yet observed. The two examples presented in
			the article, Andromeda and Sagittarius A*, are
			marked by the symbols A and S. }
		\label{boostMassDist}
	\end{figure}
	
	In order to evaluate a large class of astrophysical black
	holes we show in Fig.~\ref{boostMassDist} the DM density
	{$\rho_{\mathrm{out}}$} depending on the mass of the black
	hole and {$r_{\mathrm{out}}$}. As a second specific example we
	choose Sagittarius A* and evaluate the boost function at a
	distance to the black hole of $10\,$pc as shown in
	Fig.~\ref{boostMassDist}. Choosing the target location this
	close to the black hole would open up new possibilities for
	indirect DM detection close to black holes by observing
	annihilation or decay products from this region. However, as
	seen in Fig.~\ref{boostMassDist} the boost factor for
	Sagittarius A* is negligible and therefore it is not
	worthwhile to examine this possibility further. For other
	black holes, as seen in Fig.~\ref{boostMassDist} there is an
	optimal boost function for smaller distances and masses in the
	range of $10^8\,M_\odot$ to $10^9\,M_\odot$. However, since
	these DM overdensities are orders of magnitude too small to be
	relevant for any indirect DM observations, we conclude that,
	within the scope of our investigation, DM beams are generated
	by spinning black holes but are so faint that they do not
	significantly contribute to DM overdensities.

	\section{Discussion and conclusion}\label{sec:conclusion}
	The aim of this paper was to investigate whether beams of DM
	particles can be formed along the rotation axis of
	supermassive rotating black holes through the Penrose
	process. We showed that this is indeed the case and further
	analyzed the intensity and profile of such jets. In
	particular, they are shown to be very collimated and
	long-ranged. However, the density of the DM particles
	inside the beams is very small and cannot possibly generate
	overdensities near the Earth. Therefore, within the scope of our analysis, this mechanism
	can be excluded as a possible source of a boost for the local
	DM detection rate.
	
	We showed that the density is larger closer to the black hole
	and becomes smaller and smaller further away. Furthermore, we
	showed that a black hole mass of approximately
	{$10^{8}-10^{9}\,M_{\odot{}}$} leads to the densest beams. The
	density falls off sharply further away from the rotational
	axis, indicating that the DM jet is highly collimated. By
	tuning appropriately all parameters, a maximal value of
	approximately {$40\,\rho_{0}$} can be obtained. Furthermore,
	increasing the black hole's angular momentum creates a more
	collimated beam.
	
	We hope that our study inspires future calculations. Moreover,
	a closer analysis of the produced overdensites near the
	rotating black hole at the center of the Milky Way might allow
	to find a source of DM annihilation which is relevant for
	local DM detection.

	\begin{acknowledgments}
		We appreciate financial support of our work by the DFG cluster of
		excellence ``Origin and Structure of the Universe''. Furthermore, we
		are thankful for helpful input from Alexis Kassiteridis and comments
		from Cecilia Giavoni and Marc Schneider.
		
	\end{acknowledgments}
	
	\bibliographystyle{unsrt}
	\bibliography{DMjetBib}

\begin{thebibliography}{10}

\bibitem{Blandford:1977ds}
R.~D. Blandford and R.~L. Znajek.
\newblock {Electromagnetic extractions of energy from Kerr black holes}.
\newblock {\em Mon. Not. Roy. Astron. Soc.}, 179:433--456, 1977.

\bibitem{Penrose:1971uk}
R.~Penrose and R.~M. Floyd.
\newblock {Extraction of rotational energy from a black hole}.
\newblock {\em Nature}, 229:177--179, 1971.

\bibitem{Gariel:2010}
J.~Gariel, M.~A.~H. MacCallum, and N.~O. Santos.
\newblock {Kerr geodesics, the Penrose process and jet collimation by a black
  hole}.
\newblock 2010.

\bibitem{Adriani:2008zr}
Oscar Adriani et~al.
\newblock {An anomalous positron abundance in cosmic rays with energies 1.5-100
  GeV}.
\newblock {\em Nature}, 458:607--609, 2009.

\bibitem{Aguilar:2013qda}
M.~Aguilar et~al.
\newblock {First Result from the Alpha Magnetic Spectrometer on the
  International Space Station: Precision Measurement of the Positron Fraction
  in Primary Cosmic Rays of 0.5–350 GeV}.
\newblock {\em Phys. Rev. Lett.}, 110:141102, 2013.

\bibitem{Ambrosi:2017wek}
G.~Ambrosi et~al.
\newblock {Direct detection of a break in the teraelectronvolt cosmic-ray
  spectrum of electrons and positrons}.
\newblock 2017.

\bibitem{Aharonian:2004wa}
F.~Aharonian et~al.
\newblock {Very high-energy gamma rays from the direction of Sagittarius A*}.
\newblock {\em Astron. Astrophys.}, 425:L13--L17, 2004.

\bibitem{Weniger:2012tx}
Christoph Weniger.
\newblock {A Tentative Gamma-Ray Line from Dark Matter Annihilation at the
  Fermi Large Area Telescope}.
\newblock {\em JCAP}, 1208:007, 2012.

\bibitem{Bergstrom:2005ss}
Lars Bergstrom, Torsten Bringmann, Martin Eriksson, and Michael Gustafsson.
\newblock {Gamma rays from heavy neutralino dark matter}.
\newblock {\em Phys. Rev. Lett.}, 95:241301, 2005.

\bibitem{Bergstrom:2004cy}
Lars Bergstrom, Torsten Bringmann, Martin Eriksson, and Michael Gustafsson.
\newblock {Gamma rays from Kaluza-Klein dark matter}.
\newblock {\em Phys. Rev. Lett.}, 94:131301, 2005.

\bibitem{Hooper:2008kv}
Dan Hooper, Albert Stebbins, and Kathryn~M. Zurek.
\newblock {Excesses in cosmic ray positron and electron spectra from a nearby
  clump of neutralino dark matter}.
\newblock {\em Phys. Rev.}, D79:103513, 2009.

\bibitem{Meade:2009iu}
Patrick Meade, Michele Papucci, Alessandro Strumia, and Tomer Volansky.
\newblock {Dark Matter Interpretations of the e+- Excesses after FERMI}.
\newblock {\em Nucl. Phys.}, B831:178--203, 2010.

\bibitem{Bergstrom:2009fa}
Lars Bergstrom, Joakim Edsjo, and Gabrijela Zaharijas.
\newblock {Dark matter interpretation of recent electron and positron data}.
\newblock {\em Phys. Rev. Lett.}, 103:031103, 2009.

\bibitem{Lavalle:2007apj}
J.~Lavalle, Q.~Yuan, D.~Maurin, and X.~J. Bi.
\newblock {Full Calculation of Clumpiness Boost factors for Antimatter Cosmic
  Rays in the light of Lambda-CDM N-body simulation results. Abandoning hope in
  clumpiness enhancement?}
\newblock {\em Astron. Astrophys.}, 479:427--452, 2008.

\bibitem{Hisano:2003ec}
Junji Hisano, Shigeki Matsumoto, and Mihoko~M. Nojiri.
\newblock {Explosive dark matter annihilation}.
\newblock {\em Phys. Rev. Lett.}, 92:031303, 2004.

\bibitem{Bergstrom:2009ib}
Lars Bergstrom.
\newblock {Dark Matter Candidates}.
\newblock {\em New J. Phys.}, 11:105006, 2009.

\bibitem{Bergstrom:2012fi}
Lars Bergstrom.
\newblock {Dark Matter Evidence, Particle Physics Candidates and Detection
  Methods}.
\newblock {\em Annalen Phys.}, 524:479--496, 2012.

\bibitem{Boshkayev:2018sbj}
Kuantay Boshkayev and Daniele Malafarina.
\newblock {A model for a dark matter core at the Galactic Centre}.
\newblock {\em Mon. Not. Roy. Astron. Soc.}, 484(3):3325--3333, 2019.

\bibitem{Brandt:2010ts}
W.~N. Brandt and D.~M. Alexander.
\newblock {Supermassive Black-Hole Growth Over Cosmic Time: Active Galaxy
  Demography, Physics, and Ecology from Chandra Surveys}.
\newblock {\em Proc. Nat. Acad. Sci.}, 107:7184, 2010.

\bibitem{Tulin:2017ara}
Sean Tulin and Hai-Bo Yu.
\newblock {Dark Matter Self-interactions and Small Scale Structure}.
\newblock {\em Phys. Rept.}, 730:1--57, 2018.

\bibitem{Dolan:2011xt}
Brian~P. Dolan.
\newblock {Pressure and volume in the first law of black hole thermodynamics}.
\newblock {\em Class. Quant. Grav.}, 28:235017, 2011.

\bibitem{code}
Ottavia Balducci, Stefan Hofmann, and Maximilian Koegler.
\newblock Dark matter jets of rotating black holes,
  https://doi.org/10.5281/zenodo.6618646, June 2022.

\end{thebibliography}
	
\end{document}